# Connectivity dependence of Fano resonances in single molecules

Ali K. Ismael[1,2] Iain Grace[1] and Colin J. Lambert[1,a]

[1]*Department of Physics, Lancaster University, Lancaster, United Kingdom*

[2]*Department of Physics, College of Education for Pure Science, Tikrit University, Tikrit, Iraq*

Using a first principles approach combined with analysis of heuristic tight-binding models, we examine the connectivity dependence of two forms of quantum interference in single molecules. Based on general arguments, Fano resonances are shown to be insensitive to connectivity, while Mach-Zehnder-type interference features are shown to be connectivity dependent. This behaviour is found to occur in molecular wires containing anthraquinone units, in which the pendant carbonyl groups create Fano resonances, which coexist with multiple-path quantum interference features.

## I. INTRODUCTION

Fano resonances[1] occur when a localised state interacts with a continuum[2]. In the context of single-molecule electronics, the continuum is provided by electrodes, which feed electrons into the molecule over a continuous range of energies. The appearance of Fano resonances in electron transport through single molecules was recognised in a series of studies of a family of rigid molecules[3-5] which contained pendant groups. Subsequently it was shown that such groups could be utilised to control transport properties though single molecules[6] and enhance their thermoelectric properties[7]. The aim of the present paper is to examine how Fano resonances can be distinguished from other quantum interference (QI) effects in molecules[8-21]. This issue is of interest, because recently unequivocal realisations of room temperature QI have been demonstrated in the laboratory[22-35] and if QI is to be utilised in the design and optimisation of molecular-scale devices[36-39] and thin films, the signatures of distinct forms of QI need to be clarified.

The aim of the present work is to demonstrate that connectivity can be used to distinguish between Fano resonances and multiple-path Mach-Zehnder-like interference effects. To illustrate how connectivity can be utilised, we start with an introductory discussion of the properties of a series of tight-binding lattices, which have also been used to study Fano resonances in quantum dot structures[40-42]. Consider first the doubly-infinite one-dimensional tight-binding chain shown in Figure 1a, described by the Schrodinger equation

$$\varepsilon_0 \varphi_j - \gamma \varphi_{j-1} - \gamma \varphi_{j+1}$$
$$= E\varphi_j \quad (-\infty < j < \infty), \quad (1)$$

whose solutions are $\varphi_j = e^{ikj}$ where $-\pi < k < \pi$. Substituting this into equation (1) yields the dispersion relation $E = \varepsilon_0 - 2\gamma \cos k$. We are interested in solving scattering problems for a given choice of $E$, for which the dimensionless wave vector $k$ satisfies $k(E) = cos^{-1}[(\varepsilon_0 - E)/2\gamma]$ and the group velocity $v(E)$ (in units of the atomic spacing) is given by $\hbar v(E) = \frac{dE}{dk} = 2\gamma \sin k(E)$, where $\hbar$ is Planck's constant. In what

---

a) **Author to whom correspondence should be addressed.**
Electronic mail: c.lambert@lancaster.ac.uk

follows, the sign of $k(E)$ is chosen such that the retarded Greens function for such a chain is

$$g_{jl} = \frac{e^{ik(E)|j-l|}}{i\hbar v(E)} \quad, \tag{2}$$

which means that: (a) when $k(E)$ is real, $k(E)$ is chosen such that $v(E) > 0$ (b) when $k(E)$ is complex, $k(E)$ is chosen such that $\text{Im } k(E) > 0$. These choices ensure that relative to the source $l$, when considered as a function of $j$, $g_{jl}$ is either outgoing or decaying.

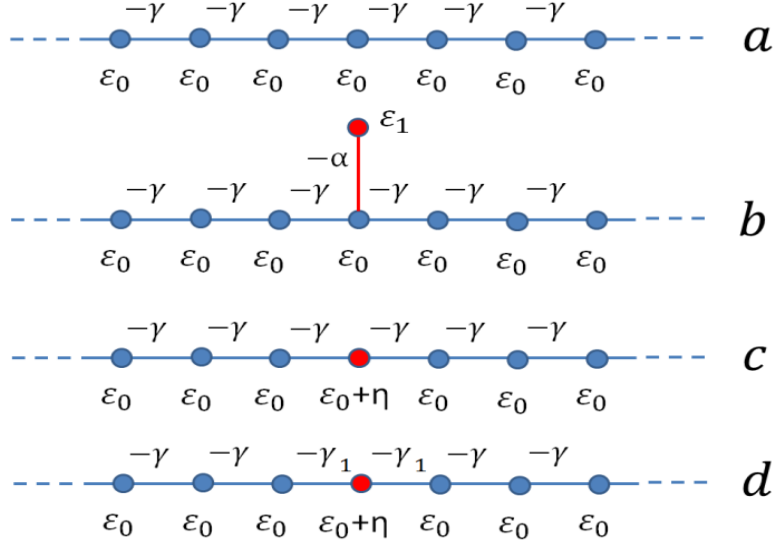

FIG. 1. A doubly-infinite one-dimensional crystalline tight-binding (TB) chain. (b) As for (a), but with a pendant site of energy $\varepsilon_1$, attached to the chain by a coupling $-\alpha$. (c) Decimating the pendant site from (b) yields an equivalent lattice with a single impurity site of energy $\varepsilon_0 + \eta$. (d) As for (c), except the impurity is coupled to the lattice by weak links $-\gamma_1$.

Now consider the chain of Figure 1b, which contains a pendant site of energy $\varepsilon_1$, connected to site $j = 0$ by a matrix element $-\alpha$. Since this pendant site interacts with the continuum of energies associated with the doubly-infinite chain, we expect a destructive Fano resonance to appear. This is most easily demonstrated by decimating the pendant site to yield the mathematically equivalent lattice shown in Figure 1c, which contains a single impurity on site $j=0$ of energy $\varepsilon_0 + \eta$, where $\eta = \alpha^2/(E - \varepsilon_1)$. Clearly for such an impurity-containing lattice, if $\eta = \infty$, the chain divides into two sections separated by an infinite tunnel barrier and therefore the transmission coefficient $T(E)$ vanishes. This occurs when $E = \varepsilon_1$ and therefore we conclude that $T(\varepsilon_1) = 0$. More generally, the energy dependence of $T(E)$ is obtained by computing the Greens function $G_{jl}(E)$ of the chain of Figure 1c and noting that for any $j \geq 0$ and $l \leq 0$,

$$T(E) = [\hbar v(E)]^2 |G_{jl}(E)|^2, \tag{3}$$

Dysons equation yields

$$G_{jl}(E) = g_{jl} + \frac{g_{j0}\eta g_{0l}}{1 - \eta g_{00}} =$$

$$\frac{g_{jl} + \eta[g_{j0}g_{0l} - g_{00}g_{jl}]}{1 - \eta g_{00}} = \frac{g_{jl}}{1 - \eta g_{00}}, \tag{4}$$

where the last step makes use of the fact that for $j \geq 0$ and $l \leq 0$, $g_{j0}g_{0l} = \frac{e^{ik(E)(j-l)}}{[i\hbar v(E)]^2} = g_{00}g_{jl}$ .
Hence, we obtain three equivalent expressions for $T(E)$, all of which vanish when $E = \varepsilon_1$ and $\eta = \infty$:

$$T(E) = \frac{1}{1 + [\eta/\hbar v(E)]^2} = \frac{4\Gamma^2}{\eta^2 + 4\Gamma^2} =$$

$$\frac{4\Gamma^2}{[E - \bar{\varepsilon}_0 + \sigma]^2 + 4\Gamma^2} \tag{5}$$

where in the last expression, $\Gamma = \hbar v(E)/2 = \gamma \sin k$ , $\sigma = 2\gamma \cos k$ and $\bar{\varepsilon}_0 = \varepsilon_0 + \eta$.

In the literature[43], it has been suggested that although the above anti-resonance can be classified as a Fano resonance, because the continuum of states within the periodic chain interferes with the state localized on the pendant site, the analogy does not hold for single molecule junctions where the continuum of the electrode states is interrupted by a junction, which acts as a tunnelling barrier. We see no compelling reason to make such a distinction, because the last expression in equation (4) holds even when tunnelling barriers are present. For example for the lattice shown in Figure 1d, where the site at $j = 0$ is connected to the 1-d periodic chains on the left and right by weak coupling elements $-\gamma_1$, which act as tunnel barriers, equation (4) still holds, but $\sigma$ and $\Gamma$ are replaced by $\Gamma = \frac{\gamma_1^2}{\gamma}\sin k$ and $\sigma = 2\frac{\gamma_1^2}{\gamma}\cos k$. Therefore as $\gamma_1/\gamma$ varies smoothly from zero to unity, the same functional form persists.

Nevertheless it is of interest to enquire how the above Fano resonances can be distinguished from other forms of destructive interference. In what follows, we propose that the key distinguishing feature lies in their connectivity dependence.

To clarify the role of connectivity consider the tight-binding ring of $N$ sites shown in Figure 2a, described by the Schrodinger equation

$$\varepsilon_0\varphi_j - \gamma\varphi_{j-1} - \gamma\varphi_{j+1} = E\varphi_j \quad (1 < j < N)$$
$$\varepsilon_0\varphi_1 - \gamma\varphi_N - \gamma\varphi_2 = E\varphi_1$$
$$\varepsilon_0\varphi_N - \gamma\varphi_{N-1} - \gamma\varphi_1 = E\varphi_N$$

The solutions to this are $\varphi_j^n = e^{i2n\pi j/N}$ where $n = 1, 2, \ldots N$ and the eigenvalues are $E_n = \varepsilon_0 - 2\gamma\cos 2n\pi/N$. Similarly, the Green's function of this closed ring[44] is

$$G_{ring}(j, i) = A\cos k(E)\left(|j - i| - \frac{N}{2}\right), \quad (6)$$

where $A = \frac{1}{2\gamma\sin k(E)\sin k(E)N/2}$. Since the Green's function of a closed system diverges when $E$

coincides with an eigenvalue, as expected, $A$ diverges when $\frac{k(E)N}{2} = n\pi$; ie when $E = E_n$.

Now consider coupling such a ring via sites $i$ and $j$ to 1-d semi-infinite crystalline leads by weak coupling elements $-\gamma_1$ and $-\gamma_2$ respectively. As noted in[44], the transmission coefficient for such a structure is

$$T(E) = \left[2\frac{\gamma_1^2}{\gamma}\sin k(E)\right]^2 |G_{ji}(E)|^2, \quad (7)$$

where $G_{ji}(E) = \frac{G_{ring}(j,i)}{1+x}$. In this expression,
$$x = \left\{1 + G_{ring}(i, i)\frac{\gamma_1^2}{\gamma}e^{ik} + G_{ring}(j, j)\frac{\gamma_2^2}{\gamma}e^{ik} + \left[G_{ring}(i, i)G_{ring}(j, j) - G_{ring}(i, j)G_{ring}(j, i)\right]\frac{\gamma_1^2}{\gamma}\frac{\gamma_2^2}{\gamma}e^{2ik}\right\}, \quad (8)$$

Therefore provided $E \neq E_n$, (so that $G_{ring}$ does not diverge) $x$ vanishes in the limit $\frac{\gamma_1}{\gamma} \to 0$ and $\frac{\gamma_2}{\gamma} \to 0$. Consequently in this weak coupling limit,

$$T(E) \approx \left[2\frac{\gamma_1^2}{\gamma}\sin k(E)\right]^2 |G_{ring}(j,i)|^2, \quad (9)$$

Equations (7) and (9) illustrate the connectivity dependence of multiple-path QI in such rings. For example when electrons enter a 6-membered ring with energy in the middle of the HOMO-LUMO gap, $E = \varepsilon_0$ and $k(\varepsilon_0) = \pi/2$, As shown in Figure 2c, for N=6, the transmission in the case of para coupling is determined by the Greens function $G_{ring}(4,1) = A$, which is non-zero and corresponds to constructive QI, whereas in the case of meta coupling (Figure 2d), it is determined by $G_{ring}(3,1) = A\cos k$, which vanishes at the gap centre and corresponds to destructive QI.

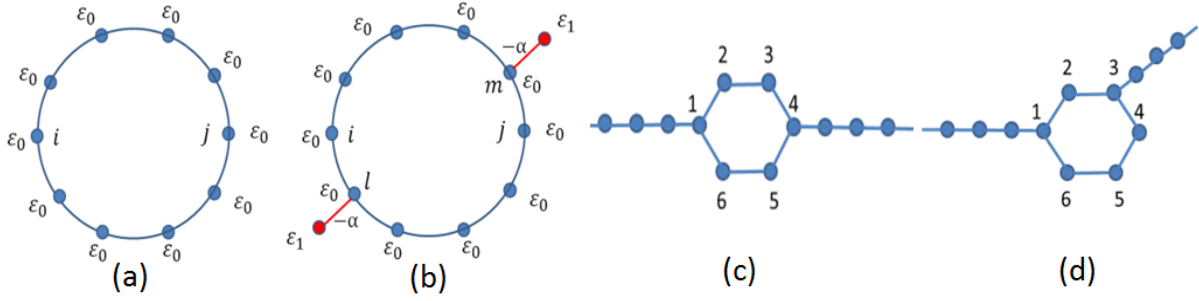

FIG. 2. (a) A TB ring of N sites. (b) A TB ring with pendant sites attached to sites l and m. (c) A para-connected 6-membered ring. (d) A meta-connected 6-membered ring.

The above sensitivity to connectivity ($i$, $j$) arises from the presence of Mach-Zehnder-type multiple-path interference in the ring, which for certain values of $i$, $j$ and $E$ can cause $G_{ring}(j, i)$ to vanish. As illustrated by the difference between the transmission coefficients of meta- and para-connected rings, this type of QI changes from destructive to constructive by only a single site shift of the connecting site (ie by changing the connection from site 3 to 4). We now examine the effect of introducing pendant sites into such rings, as illustrated by Figure 2b, where the pendant sites of energy $\varepsilon_1$ are coupled to sites $l$ and $m$ of the ring. After decimation, this structure again maps onto an equivalent lattice with impurity sites located at positions $l$ amd $m$, of energy $\varepsilon_0 + \eta$, where $\eta = \alpha^2/(E - \varepsilon_1)$ and when $E = \varepsilon_1$, the structure divides into two halves. Consequently $T(\varepsilon_1)$ vanishes, provided the sites $i$ and $j$, which connect the ring to external crystalline leads, become disconnected by divergent sites at $l$ and $m$. Provided the latter condition is satisfied, changing the connectivity does not change the energy $E = \varepsilon_1$ at which the Fano destructive interference occurs. On the other hand for $E \neq \varepsilon_1$, transmission remains finite and Mach-Zehnder interference features remain, albeit with energies shifted by the presence of the pendant sites.

## II. TRANSPORT CALCULATIONS

After the above discussion of the main concept, in this section, we first employ a tight-binding model to probe the connectivity dependence of transport through the tight binding lattices a-f of Figures 3 and 5 and then use density functional theory (DFT) to probe transport through the anthraquinone-based molecules, which are realisations of the abstract structure of Figure 2b. As expected, we find that they possess connectivity-independent Fano resonances, which coexist with connectivity-dependent Mach-Zehnder-type interference features.

For the lattice in Figure 3, each atom within the core is assigned a site energy $\varepsilon_0 = 0$ and a nearest neighbour hopping element $-1$. The atoms i, j are attached to the terminal sites of semi-infinite 1-d crystalline leads by a hopping element $-\gamma_1$. In what follows, we choose $\gamma_1 = -0.2$. The pendant site is assigned a site energy $\varepsilon_1 = -0.22$ and is coupled to its nearest neighbour by a coupling $\alpha = -0.1$.

The structures in Figures 3 are examples of bipartite lattices, in which sites can be numbered such that odd numbered sites connect to even numbered sites only and vice versa. Consequently if the sites i, j attached to the leads are both even or both odd (as in Figure 3a), multiple-path destructive interference is expected at E=0. On the other hand, this destructive QI at E=0 is expected to disappear when one of the connection points is shifted by one site, such that i is odd and j is even or vice versa, as in Figure 3b. This connectivity dependence of multi-path interference dips is clearly evident in the transmission curves of Figure 4. On the other hand, the transmission curves also show destructive QI at $E = \varepsilon_1$, arising from Fano resonances associated with the pendant groups and these persist for both connectivities.

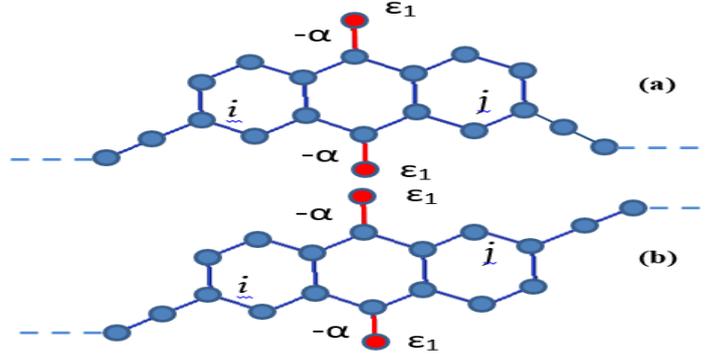

FIG. 3. (a) A bipartite lattice with odd to odd connectivity (b) A bipartite lattice with odd to even connectivity.

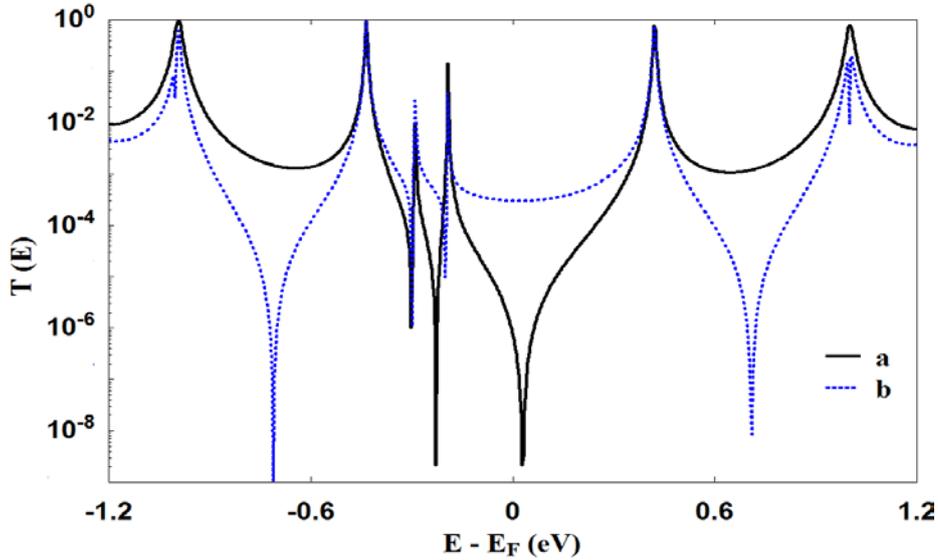

FIG. 4. Transmission curves for lattices a and b of figure 3. ($\varepsilon_1 = -0.22$)

As noted in the previous section, pendant groups can be decimated to yield equivalent lattices with impurity site, whose site energies diverge when $E = \varepsilon_1$. In the lattices of Figure 3, this divergence cuts the structures into separate halves and therefore T(E) vanishes at $E = \varepsilon_1$, for both connectivities. In contrast for lattices (c) and (d) of Figure 5, only the lower section of the central ring is bisected and a current path from i to j persists. Therefore in this case the transmission does not vanish at $E = \varepsilon_1$. For comparison, Figure 6 shows the transmission coefficients of lattices (e) and (f), which contain no pendant groups. Since these are bipartite, (e) shows a transmission dips at E=0 due to its odd-odd connectivity, whereas (f) shows no such dip, due to its odd-even connectivity.

Comparison with the transmission curves of (c) and (d) show that these connectivity-dependent features persist in the presence of pendant groups and furthermore the presence of such groups leads to additional Fano-like interference features. However since these do not bisect the lattice, an anti-resonance does not occur at $E = \varepsilon_1$. We therefore refer to this as an "incomplete Fano resonance". Instead of a connectivity-independent anti-resonance at $E = \varepsilon_1$, multiple-path interference features occur in the vicinity of $E = \varepsilon_1$, whose energetic location is connectivity dependent.

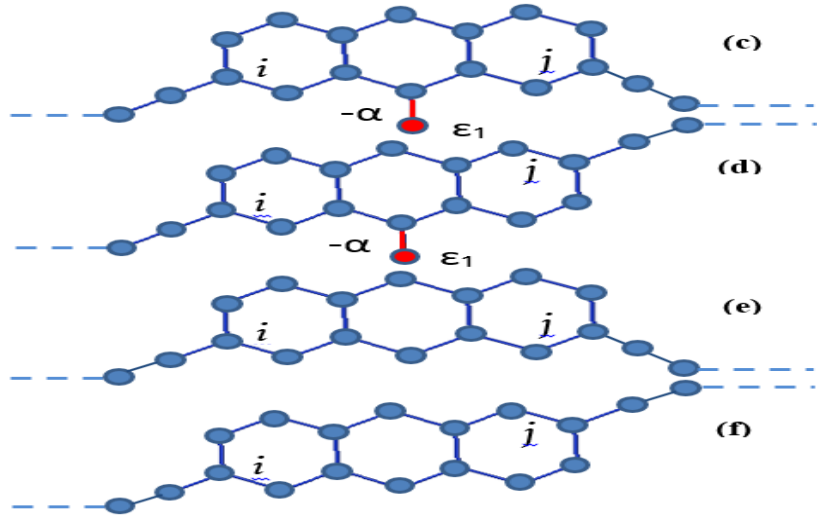

FIG. 5. (Upper two figures) Two lattices (c) and (d) with pendant sites, but different connectivities to the leads. (Lower two figures) Two lattices (e) and (f) with different connectivities to the leads and no pendant sites.

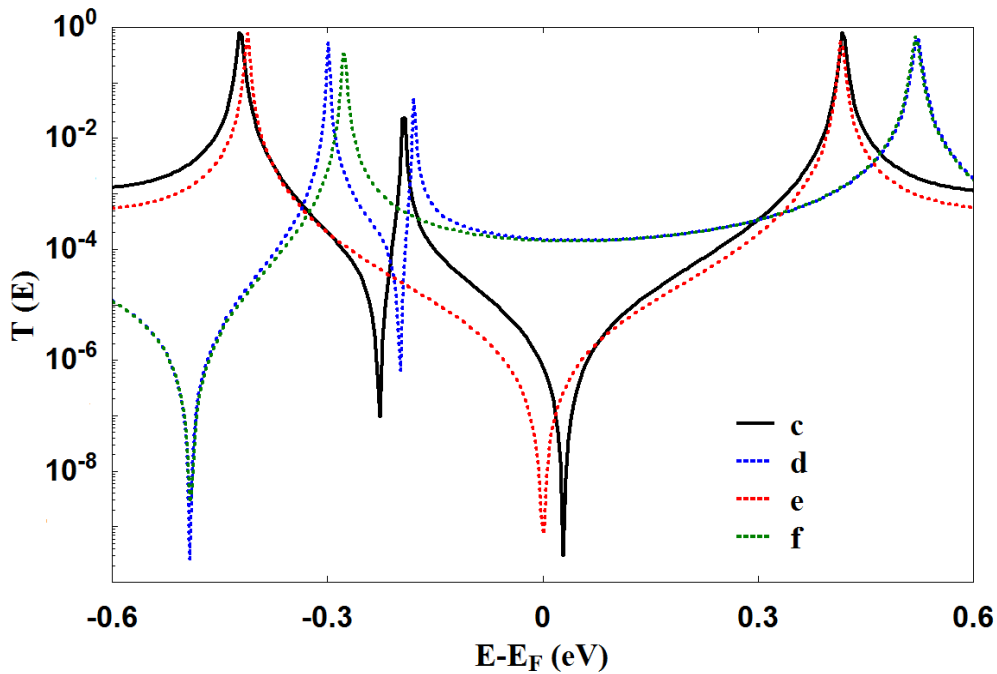

FIG. 6. Transmission curves for the four lattices (c-f) of figure 5.

## III. DFT CALCULATIONS

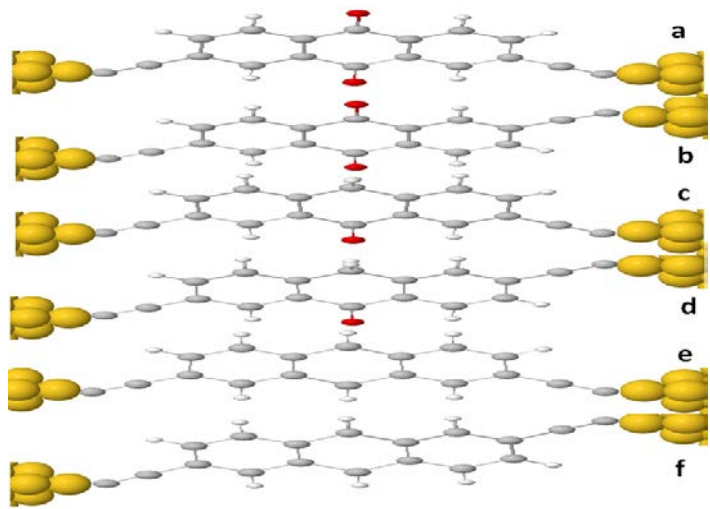

FIG. 7. (Upper two Figures) Two lattices (a) and (b) with two pendant sites, but different connectivities to the leads. (Middle two Figures) Two lattices (c) and (d) with one pendant sites, but different connectivities to the leads. (Lower two Figures) Two lattices (e) and (f) with different connectivities to the leads and no pendant sites.

So far the discussion has focused on tight-binding representations of the molecules shown in Figure 7. For molecules (a) and (b) containing two carbonyl groups, Figure 8 shows DFT results for the transmission coefficients (see methods for more details). In common with the tight binding results of figure 4, both a and b show alternating destructive interference features. In the case of a, such a feature occurs between the LUMO and LUMO+1, whereas in the case of b, they occur between the HOMO and LUMO and LUM+2 and LUMO+3. On the other hand for both molecules, destructive interference features appear near E=-2.4eV, which are almost independent of connectivity, signaling the presence of a Fano resonance.

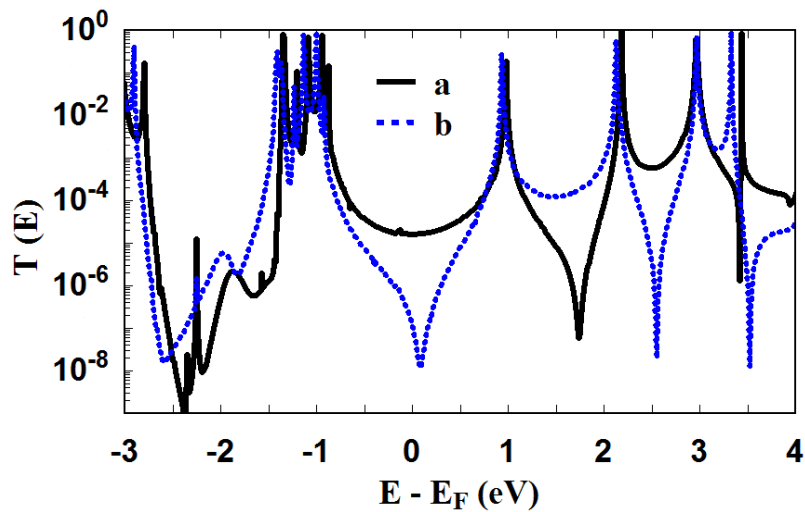

FIG. 8. DFT transmission curves for molecules a and b of figure 7. **Erratum:** Note that in the published version of this paper, the labels of curves a and b of figure 8 should be interchanged. This means that the analogue of the interference dip near E=0 in figure 4 occurs near E=1.75 eV in the solid curve of figure 8.

Figure 9 shows DFT results for molecules (c) – (f) of Figure 7. For molecules (e) and (f), which contain no pendant atoms, connectivity (e) possesses a transmission dip near E=3eV due to the odd-odd connectivity, which disappears upon switching to the odd-even connectivity of molecule (f). This transmission dip is also present for the odd-odd connectivity molecule (c) containing a single pendant group. Again this dip disappears upon switching to the odd-even connectivity of molecule (d). Furthermore, in common with the tight-binding results (c) and (d) of Figure 6 and in contrast with the transmission curves of molecules (e) and (f), molecule (c) also possesses a lower energy destructive QI feature at E=1.7 eV, which moves to lower energies upon switching connectivity to that of molecule (d), signaling the presence of an incomplete Fano resonance.

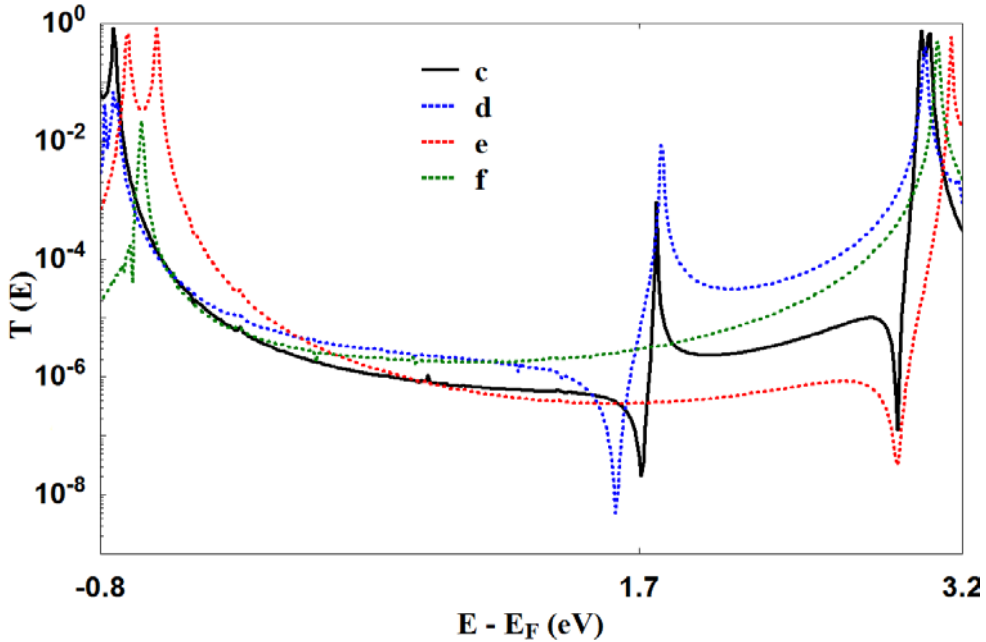

FIG. 9. DFT transmission curves for the four lattices (c-f) of figure 7.

## IV. CONCLUSION

In summary we have shown that transport through anthraquinone-based molecular wires containing one or two pendant atoms exhibit Fano resonances, which can be differentiated from multiple-path Mach-Zehnder type interference through their resilience to changes in their connectivity to electrodes. This resilience is expected on general grounds, whenever a Fano resonance causes the wave-function of electrons traversing a molecule to decouple at a certain energy. If this does not occur, the Fano resonance is incomplete and the associated interference dip becomes a connectivity-dependent multiple path interference feature. Such interference features have the potential to increase the sensitivity of single-molecule sensors and the thermoelectric performance of single molecules[6,7] and molecular films. Therefore the ability to preserving their energetic locations while selecting desirable connectivities is a useful ingredient in design strategies aimed at optimizing such properties.

## V. COMPUTATIONAL METHODS

Electronic structure calculations were performed using the DFT code SIESTA[45]. The optimum geometry of the isolated a-f molecules were obtained by relaxing the molecules until all forces on the atoms were less than 0.05 eV/Ang. The SIESTA calculations employed a double-zeta plus polarization orbital basis set, norm-conserving pseudopotentials, an energy cutoff of 200 Rydergs defined the real space grid and the exchange correlation functional was LDA. To calculate the conductance through these two molecules, they were attached to gold leads via the pyridyl anchor groups. The leads were constructed of 6 layers of (111) gold each contain 30 gold atoms and the optimum binding distance was calculated to be 2.3Å between the terminal carbon atoms and a 'top' gold atom. A Hamiltonian describing these structures were produced using SIESTA and the zero-bias transmission coefficient T(E) was calculated using the Gollum code[46].

## ACKNOWLEDGEMENTS

This work was supported by the Swiss National Science Foundation (no. 200021-147143) as well as by the European Commission (EC) FP7 ITN "MOLESCO" project no. 606728 and UK EPSRC, (grant nos. EP/K001507/1, EP/J014753/1, EP/H035818/1), and the Iraqi Ministry of Higher Education, Tikrit University (SL-20). A. K. I. acknowledges financial support from Tikrit University.